# Performance Metrics (Error Measures) in Machine Learning Regression, Forecasting and Prognostics: Properties and Typology

**Alexei Botchkarev**
Principal, GS Research & Consulting,
Adjunct Prof., Department of Computer Science, Ryerson University
Toronto, Ontario, Canada

**Abstract**

Performance metrics (error measures) are vital components of the evaluation frameworks in various fields. The intention of this study was to overview of a variety of performance metrics and approaches to their classification. The main goal of the study was to develop a typology that will help to improve our knowledge and understanding of metrics and facilitate their selection in machine learning regression, forecasting and prognostics. Based on the analysis of the structure of numerous performance metrics, we propose a framework of metrics which includes four (4) categories: primary metrics, extended metrics, composite metrics, and hybrid sets of metrics. The paper identified three (3) key components (dimensions) that determine the structure and properties of primary metrics: method of determining point distance, method of normalization, method of aggregation of point distances over a data set. The paper proposed a new primary metrics typology designed around the key metrics components. The suggested typology has been shown to cover most of the commonly used primary metrics – total of over 40. The main contribution of this paper is in ordering knowledge of performance metrics and enhancing understanding of their structure and properties by proposing a new typology, generic primary metrics mathematic formula and a visualization chart.

Note: This is a draft paper submitted to a peer-reviewed journal.

## Introduction

Performance evaluation is an interdisciplinary research problem. Performance metrics (error measures) are vital components of the evaluation frameworks in various fields. In machine learning regression experiments, performance metrics are used to compare the trained model predictions with the actual (observed) data from the testing data set (e.g. Makridakis, Spiliotis and Assimakopoulos, 2018; Botchkarev, 2018a). Forecasting has a long history of employing performance metrics to measure how much forecasts deviate from observations in order to assess quality and choose forecasting methods, especially in support of supply chain or predicting workload for software development (e.g. Carbone and Armstrong, 1982; De Gooijer and Hyndman, 2006). Prognostics - an emerging concept in condition-based maintenance (CBM) of critical systems in aerospace, nuclear, medicine, etc. – heavily relies on performance metrics (e.g. Saxena et al, 2008). In a generic sense, performance metrics are linked to the concepts of distance and similarity. Deza and Deza (2016) state that "similarity measures are needed in almost all knowledge disciplines."

Classification is one of the main topics of scientific research (Parrochia, n.d.). Each knowledge domain, as a subject of scientific research, requires classification systems (typology) to structure the contents in a systematic manner. Categories of the typology are defined based on resemblances (or differences) of items/objects in a specific context. Typologies are helpful in ordering and organizing knowledge, defining the scope and simplifying studies, facilitating information retrieval and detecting duplicative objects

The intention of this paper was to review existing performance metrics classifications and develop a typology that will help to improve our knowledge and understanding of a variety of metrics and facilitate their use in machine learning regression, forecasting and prognostics.

The rest of the paper is structured as follows. First, we describe methodology of the study. Then, we provide a literature review. In the next section, we describe a proposed metrics framework, which includes the following categories: primary metrics, extended metrics, composite metrics and hybrid sets of metrics. The main attention and space in this paper is focused on the properties and typology of the primary metrics. The final sections present discussion and concluding remarks.

## Methodology

**Objectives.** The first objective of this study was to provide an overview of a variety of the performance metrics and approaches to their classification (grouping/systematization). The main goal of the study was to develop a new typology that would contribute to ordering knowledge of performance metrics and enhancing understanding of their structure and properties.

**Method.** Several research methodologies were used to achieve the objectives: identification of related peer-reviewed papers, critical literature review, critical thinking and inductive reasoning. The search was conducted in Google Scholar and several databases through the EBSCO integrated search including Health Business Elite, Health Policy Reference Center, BioMed Central, Business Source Complete, MEDLINE Complete, CINAHL Complete, PubMed, The Cochrane Library, etc. Around 500 papers where retrieved and previewed. Over 80 papers where selected, reviewed in more detail and cited in the paper.

**Terminology and abbreviations.** As this paper covers research in an interdisciplinary area, terminology may vary from field to field.

Our main focus is on performance metrics. In literature, many terms are used with close meaning, e.g. measure, distance, similarity, dissimilarity, index, etc.

Different terms are used in literature regarding grouping performance metrics, e.g. classification, taxonomy, etc. In the literature review, we use the terms used by the authors of the papers under consideration. Later in the paper, we refer to our construct as typology.

Multiple performance metrics are considered in the paper. Commonly, we refer to them using abbreviations. A list of all metrics abbreviations mentioned in the paper is provided in Appendix 1. Usually, the first letters in abbreviations use: M for mean (arithmetic), Md for median, GM for geometric mean.

Mathematical definitions of performance metrics are shown in Appendix 2. These metrics are implemented in R Studio (e.g. packages MLmetrics, forecast) and in Azure Machine Learning Studio (e.g. Botchkarev, 2018b). Some metrics have alternative definitions. They are listed in Appendix 3.

Performance metrics are designed to compare two data sets. We refer to them as actual, $A = (A_1, A_2, \ldots, A_j)$ and predicted, $P = (P_1, P_2, \ldots, P_j)$. In literature, depending on the research field, actual may be referred to as observed or measured, and predicted may be called forecasted, modeled, simulated, estimated.

**Literature Review**

A large variety of metrics has been suggested and used in many knowledge areas. In 1995, Makridakis and Hibon boldly stated that "there are fourteen accuracy measures which can be identified in the forecasting literature". It seems that no other author risked offering an exhaustive list of metrics. Usually, a list of metrics is accompanied with qualifiers: most popular, commonly, widely or frequently used, etc. There are many analytic reviews covering dozens of metrics. Kyriakidis, et al (2015) studied 24 metrics used in air quality forecasting. De Gooijer and Hyndman (2006), in a review covering 25 years of time series forecasting, list 17 commonly used accuracy measures. Shcherbakov et al (2013) presented a survey of more than twenty forecast error measures. Prasath et al (2017) studied 54 (fifty-four) measures and their effect on machine learning of K-Nearest Neighbor Classifier (KNN). Numerous distance metrics from diverse knowledge domains are compiled and briefly described in the Encyclopedia of distances (Deza and Deza, 2016).

Some metrics are more popular than the others. Several researchers conducted surveys of organizations and practitioners to understand the frequency of use or importance of different metrics. A variety of metrics were identified in these surveys. However, top most common metrics came up in many studies. Table 1 shows three metrics found most popular in the independent surveys that were performed over a timeline of 25 years: mean square error (MSE) (or root MSE (RMSE)), mean absolute error (MAE) and mean absolute percentage error (MAPE).

**Table 1. Top three metrics identified in the surveys, percentage**

| Metrics | C&A, 1982 | M&K, 1995 | M et al, 2006 | F&G, 2007 |
|---|---|---|---|---|
| Mean square error (MSE) or Root MSE (RMSE) | 34 | 10 | 6 | 9 |
| Mean absolute error (MAE) | 18 | 25 | 20 | 36 |
| Mean absolute percentage error (MAPE) | 15 | 52 | 45 | 44 |

**Note:**
C&A, 1982: (Carbone and Armstrong,1982); M&K, 1995: (Mentzer and Kahn, 1995); M et al, 2006: (McCarthy et al., 2006); F&G, 2007: (Fildes and Goodwin, 2007).

Data in the Table 1 reveals that preferences towards metrics have changed over the years. In the 1980s, the prevalence of the MSE/RMSE was quite clear with 34 percent – almost twice as high as of the other two metrics. However, in the 1990s, MAPE moved in the leading position and kept it in the 2000s with over 40 percent. MAE retains the second place in all surveys. It should be noted that surveys illustrated in Table 1 were conducted using different methodologies (e.g. types of respondents, sample sizes, acceptance of multiple selections, etc.). So, the comparative results should be treated as qualitative trends rather than exact numbers.

Even most popular metrics have been scrutinized from time to time and strongly criticized or even rejected. Here are some examples.

Armstrong and Collopy (1992) stated that RMSE (arguably one of the top-used metrics) was not reliable, and was inappropriate for comparing accuracy across time series. Later, Willmott and Matsuura (2005) and Willmott, Matsuura and Robeson (2009) found that RMSE has "disturbing characteristics" and is inappropriate for use as an error measure. The authors extended their conclusion on all square error measures (e.g. standard error). They recommended RMSE not to be reported in the literature and strongly advised in favour of using MAE. Chai and Draxler (2014) disputed these conclusions, at least partially, and presented arguments against avoiding RMSE.

Makridakis (1993) criticized the use of RAE as not meaningful for decision making.

Foss et al (2003) concluded that MMRE (MAPE), another very popular metric, is "an unreliable selection criterion and may have misled the entire software engineering discipline." Still, according to a number of surveys reviewed by Gneiting (2011), MAPE is the most commonly used measure for assessing forecasts in organisations.

Li (2017) asserted that correlation coefficient (R) and the coefficient of determination ($R^2$) should not be used as measures to assess the accuracy of predictive models for numerical data (because they are biased, insufficient or misleading).

Discussions on which metric to use are common in the literature. Usually, they are based on the premise that there could be a single "ideal" metric that beats all others in all situations. Paradoxically, a drive for having a best single metric, leads to an opposite result – the number of metrics tend to increase steeply.

New metrics are being developed and published regularly. For example, recent paper titles introducing new measures include: "Novel metrics…" (Grigsby et al, 2018), "A better measure…" (Tofallis, 2015), "A new accuracy measure…" (Chen, Twycross and Garibaldi, 2017), "A new metric…" (Kim and Kim, 2016), "New Statistical Indices…" (Kyriakidis, et al, 2015), "New accuracy measures…" (Bratu, 2013), "…new methods for measuring forecast error" (Mathai, et al, 2016). Two approaches are commonly used to develop new metrics. First is focused on modifying existing measures to adjust them to task-specific conditions (e.g. Grigsby et al, 2018; Bratu, 2013; Monero et al, 2013; Mathai, et al, 2016). The second approach is to combine the information contained in several existing measures (e.g. Kyriakidis, et al, 2015).

Still, no consensus on the "best" metric has been achieved. On the contrary, another notion is getting popularity. Researchers express a more practical view that there is no need to strive for a single best metric. This is an unrealistic goal - "a quest for an ideal". Silver et al (1998) pointed out that ''no single measure is universally best''. Chai and Draxler (2014) clarified that "as every statistical measure condenses a large number of data into a single value, it only provides one projection of the model errors emphasizing a certain aspect of the error characteristics of the model performance." This notion is supported by Armstrong (1985), Mahmoud (1987), Fildes and Goodwin (2007), Kyriakidis, et al, (2015), etc.

There is a foundational point which needs to be mentioned considering performance metrics. Evaluation error (deviation of actual and predicted values) is a random variable. Its complete description is possible only with probability density function or moments, if they exist.

Certain terminology clarifications are needed for better understanding of the rest of the paper.

Some popular metrics are referred to as *scale-dependant* (Hyndman, 2006) or *dimensioned* (Willmott and Matsuura, 2005) as errors have physical dimensions and expressed in the units of the data under analysis (variable of interest), e.g. MAE, RMSE. Note that the condition to categorize a metric as dimensional is two-fold: first, it must have a dimension, and, second, the dimension must be the same as of the variable of interest. For example, if we use machine learning regression to predict cost of a medical intervention, measured in dollars, then the mean absolute error will also be found in dollars. By the same token, predicting quantities with dimensions in time, speed, distance, etc. measured in dimensional units, respectively, second, mile per hour, kilometer, etc., metrics will preserve the same units.

Two caveats need to be considered.

First, certain metrics, although bear physical dimension, e.g. MSE and other squared error metrics, strictly speaking, should not be included in the dimensioned group, because their dimensions are different (changed) from the dimension of the variable of interest. For example, the cost prediction exercise mentioned above, will result in MSE measured in "squared dollars".

Second, certain variables of interest have no physical dimension, i.e. dimensionless (Dimensionless Quantity, n.d.). Examples of dimensionless quantities include: GDP ratio, coefficient of determination, elasticity, etc. (List, n.d.). Sometimes dimensionless quantities are given special names: percentages, degrees, decibels, radians, etc. Applying metrics to dimensionless variables of interest will provide dimensionless results. Paradoxically, applying MAE, RMSE metrics in these cases are still usually included in a dimensioned group. So the underline idea is that the metric should not be *changing* the nature (dimensional or dimensionless) of the input data.

By contrast, there is another group of metrics that do not have dimension and referred to as dimensionless (Dimensionless Quantity, n.d.) or scale-free, scaled, or scale-independent. Commonly, dimensionless metrics involve mathematical division of quantities of the same dimensional units (e.g. ratios, relative, percentage indicators), e.g. MAPE.

It should come as no surprise, that with a multitude of available performance metrics, research efforts are taken to organize them into categories according to common characteristics and properties for easier study, design and thoughtful application.

Makridakis and Hibon (1995) proposed a classification of error metrics by two criteria: the character of measure (absolute, relative to a base or other method, relative to the size of errors) and the type of evaluation (a single method, more than one method, in comparison to some benchmark). They presented results in a table format: character of measures as rows and types of evaluation as columns. They applied the classification to a set of 14 metrics they studied and placed metric titles in the cells of intersecting criteria. It can be seen from the table that some metrics (e.g. MAPE and MdAPE) were assigned to two cells. It reveals that the classification criteria are not mutually exclusive (overlapping) which is not good for a classification. To the best of our knowledge, this was the first attempt to build a formal error metrics typology.

Hyndman (2006) suggested classifying metrics into four (4) groups:

- scale-dependent metrics (e.g. MAE, GMAE);
- percentage-error metrics (e.g. MAPE);
- relative-error metrics (e.g. MdRAE, GMRAE);
- scale-free error metrics (e.g. MASE).

This classification is simple, intuitively clear (at least for some metrics) and has been widely used in the literature.

However, in the logical sense, this classification is not perfect – it has overlappings. It appears that the groups are categorised based on whether the metric has a scale (i.e. measured in certain units) or not. Following this logic, the classification should consist of only two top-level classes: scale-dependent and scale-free. Percentage and relative metrics should be included in the scale-free metrics. Further, percentage metrics should be a subclass of the more general relative metrics (at least linguistically, although algorithmic relationship could be more complicated).

Also, it should be noted that Hyndman (2006) includes MSE into scale-dependent group (claiming that the error is "on the same scale as the data"). This requires clarification because the MSE has a dimension of the squared scale/unit. To bring the MSE to the scale of the data we need to take a square root which results in another metric – RMSE.

Similar, but slightly different, classification was proposed by Hyndman and Koehler (2006). It acknowledged five groups:

- scale-dependent measures (e.g. MSE, RMSE, MAE, MdAE);
- measures based on percentage errors (e.g. MAPE, MdAPE, RMSPE, RMdSPE, sMAPE, sMdAPE);
- measures based on relative errors (e.g. MRAE, MdRAE, GMRAE);
- relative measures (e.g. RelMAE, CumRAE);
- scaled errors (e.g. MASE, RMSSE, MdASE).

This classification delineates relative metrics into measures based on relative individual errors and metrics based on combination of measures (dividing one metric by another).

Shcherbakov et al (2013) used a forecast error classification which is similar to Hyndman and Koehler's (2006) and included seven groups: absolute forecasting errors, measures based on percentage errors, symmetric errors, measures based on relative errors, scaled errors, relative measures and other error measures.

Cha (2007) analyzed similarity measures as they apply to the comparison of the probability density functions. He suggested a classification which included nine (9) groups:

- $L_p$ Minkowski family measures (e.g. Euclidean, City block (Manhattan), Chebyshev);
- $L_1$ family measures (e.g. Average Manhattan – otherwise referred to as mean character distance or mean absolute error or Gower, Kulczynski distance, Soergel distance). They are based on Manhattan normalized absolute difference;
- Intersection family (e.g. Wave Hedges, Czekanowski);
- Inner product family (e.g. Kumar-Hassebrook, Dice);
- Fidelity family or Squared-chord family (e.g. fidelity, Bhattacharyya);
- Squared $L_2$ family (e.g. squared Euclidean, Neyman);
- Shannon's entropy family (e.g. Kullback- Leibler, Jeffreys);
- Combinations – measures utilizing multiple approaches from previous groups;
- Vicissitude measures (e.g. Vicis-Wave Hedges, Vicis symmetric).

This publication is widely cited (over 1,200 citations as of July 2018). However, the criteria of grouping metrics into categories were not explicitly stated, and there were some inconsistencies in assigning measures to the groups. For example, generalized Minkowski measure is listed as a separate measure in the Minkowski family. Some groups include distances from other groups, e.g. Family $L_1$ includes distances from the Intersection family, and Squared $L_2$ family includes distances from the Inner Product Family.

Cha's classification has been applied in several studies. Prasath et al (2017) used Cha's (2007) classification (with the exception of the Intersect family) to study 54 (fifty four) distance and similarity measures effect on the performance of K-Nearest Neighbor Classifier. Tschopp and Hernandez-Rivera (2017) used Cha's (2007) classification to study similarity and distance measures for vector-based datasets (e.g. histograms, signals, probability distribution functions). Hernández-Rivera, Coleman and Tschopp (2016) used Sha's (2007) classification to study similarity measures in application to X-ray diffraction patterns.

Jousselme and Maupin (2012) researched dissimilarity measures within the mathematical framework of evidence theory and presented a classification and general formulations for each category of measures. Their classification includes five categories/families. Four categories are the same as in Cha's classification (2009): Minkowski, Inner product, Fidelity and Information-based (Shannon). The fifth one is a Composite family based on the notion of two combined components: one that represents a measure of structural dissimilarity and the second that measures "information change relatively to orthogonal sum".

Weller-Fahy, Borghetti and Sodemann (2015) surveyed distance and similarity measures used within network intrusion anomaly detection. They grouped distance measures into four (4) types:

- Power distances which are based on mathematical expressions involving raising to power (e.g. Euclidean, Manhattan, Mahalanobis, Heterogeneous distance);
- Distances on distribution laws (probability-related) (e.g. Bhattacharya coefficient, Jensen, Hellinger);
- Correlation similarities and distances (e.g. Spearman, Kendall, Pearson);
- Other similarities and distances which do not fit into the three main categories).

Cunningham (2009) developed a taxonomy of similarity mechanisms for case-based reasoning which includes four (4) groups:
- Direct mechanisms (e.g. Minkowski, Manhattan, Euclidean);
- Transformation-based mechanisms (e.g. Edit Distance (Levenshtein Distance), alignment measures for biological sequences, Earth Mover Distance);
- Information theoretic measures (e.g. compression-based similarity, GenComress);
- Emergent measures arising from an in-depth analysis of the data (e.g. Random Forest, Cluster Kernels).

Some authors, without attempting to build a complete taxonomy, suggest grouping metrics by certain aspects, e.g. characteristic of error measured. Morley, Brito and Welling (2018) grouped metrics by the nature of measured statistic: accuracy (e.g. MSE, RMSE, MdAE, etc.) and bias (e.g. ME, MPE, etc.).

**Performance Metrics Framework**

Based on the analysis of the structure of numerous performance metrics presented in the literature, we propose a framework of metrics: primary metrics, extended metrics, composite metrics, and hybrid sets of metrics. Outline and examples of each category follow.

**Primary metrics** is arguably the most numerous category and include commonly used metrics such as MAE, MSE, sMAPE, etc. As it is shown in the next section, the structure of the primary metrics involves three steps: calculating point distance, performing normalization and aggregating point results over a data set. Refer to the next section for detailed description and analysis. Also, these metrics are used for construction of the metrics in other categories.

**Extended metrics** are commonly based on the primary metrics with additional normalization. The delineation with primary metrics is that normalization is performed after aggregation. Examples include:

- Normalized Root Mean Squared Error: NRMSE_sd = RMSE/sd -normalized by the standard deviation of the actual data; or NRMSE_max-min =RMSE/(maxA – minA) - normalized by the difference between maximum and minimum actual data; or NRMSE_m = RMSE/$\bar{A}$ - normalized by the mean of actual data, also known as coefficient of variation of the RMSE (CVRMSE) (Aman, Simmhan and Prasanna, 2015; Aman, Simmhan and Prasanna, 2011).
- MAD/Mean ratio (Hoover, 2006; Kolassa and Schütz, 2007).

**Composite metrics** involve two or more primary metrics which are combined to produce a single result. Examples of composite metricsinclude:

- Mean Absolute Scaled Error: MASE=MAE/MAE$_{ib}$, where MAE$_{ib}$ is MAE from an in-sample naïve forecast (Hyndman and Koehler, 2006).

- Relative Mean Absolute Scaled Error: RelMAE = MAE/MAE$_b$, where MAE$_b$ is MAE from a benchmark method, e.g. Hyndman and Koehler (2006), and relative geometric root mean square error (RGRMSE) (Syntetos and Boylan, 2005).

- Relative Root Mean Squared Error: RelRMSE = RMSE/RMSE$_b$, where RMSE$_b$ is RMSE from a benchmark method, e.g. Chen, Twycross and Garibaldi (2017), Thomakos and Nikolopoulos (2015). Note that RelRMSE is also known as Theil's U or U2 (De Gooijer and Hyndman, 2006).

Syntetos and Boylan (2005) observed that metrics which have a term 'relative' in their title can be built by combining any methods and suggested to group them into 'accuracy measures relative to another methods'.

Vogt et al (2018) tested combinations of up to six metrics in the dynamic simulation of buildings energy consumption. They recommended a composite metric calculated as a sum of four equally weighted statistical indices: the Coefficient of Variation of Root Mean Square Error (CV(RMSE)), the Normalized Mean Error (NME), the standardized contingency coefficient, and the coefficient of determination.

**Hybrid sets of metrics** are represented by several metrics (two or more) which are used in the same experiment with several output results. These sets are not intended to be combined in a single mathematical structure to provide a single-number output. Not any list of metrics can constitute a hybrid set. In a hybrid set, proposed metrics should be used to deliver mutually complementary properties providing better understanding of performance errors, e.g. measuring bias and accuracy. Using hybrid sets is in line with Fildes and Goodwin's (2007) advice of using multiple forecasting accuracy measures.

Kyriakidis et al (2015) developed a set of performance indices to evaluate artificial neural network models for air quality forecasting.

Another hybrid set of metrics was introduced by Morley, Brito and Welling (2018). They proposed two new metrics to be used in conjunction in radiation belt electron flux modeling and forecasting: the median symmetric accuracy and the symmetric signed percentage bias the use.

Zhang et al (2015) were searching for a set comprehensive, consistent, and robust metrics to assess performance of solar power forecasts. They recommended a suite of metrics consisting of MBE, standard deviation, skewness, kurtosis, distribution of forecast errors, Rényi entropy, RMSE, and OVERPer.

In our view, development of the hybrid sets of metrics should be on the top of the research agenda. Items of the agenda may include: studies on informational relationships of metrics; developing recommendations on avoiding redundancy of metrics compiled into a hybrid set; exploring ways of building minimum sets of metrics sufficiently describing error performance (e.g. Tian et al, 2016).

**Primary Metrics Typology**

Analysis of multiple performance metrics used for evaluation in many fields led to identification of three (3) key components (dimensions) that determine the properties of metrics and can be used for designing typology:

− Method of determining point distance, $\mathbb{D}$.
− Method of normalization, $\mathbb{N}$.
− Method of aggregation of point distances over a data set, $\mathbb{G}$.

This approach to building a typology is usually referred to as morphological typology - a scientific method widely used in many fields, especially in linguistics, biology, astronomy, etc.

A generic formula defining a primary performance metric can be written as follows:

$$\mathbb{m} = \mathbb{G}^z_{j=1,n} \{\mathbb{N}^z[\mathbb{D}^z(A_j, P_j)]\}$$

where $A_j$ – actual value; $P_j$ – predicted value; $n$ – size of the data set; $z$ – numerical index of the method (not 'to the power of' symbol).

The meaning of the formula is in sequential determining the point distance between the actual and predicted values, normalizing it and then aggregating over a complete data set. All performance metrics explicitly contain components of determining the point distance and aggregation. Normalization component is optional, i.e. in some metrics $\mathbb{N} = 1$.

Note that to simplify notation, we are not using superscript in the individual realizations of the methods, i.e. for $z = 1$ we write $\mathbb{D}1$, not $\mathbb{D}^1$.

Table 2 demonstrates most common methods which will be described in the subsections below. The fact that each category has almost the same number of options (4-5) is just a coincidence. The list of methods in the typology is not intended to be comprehensive. Only most popular methods are included.

**Table 2 Performance metrics typology components**

| Point Distance, $\mathbb{D}$ | Normalization, $\mathbb{N}$ | Aggregation, $\mathbb{G}$ |
|---|---|---|
| Error (magnitude of error): $\mathbb{D}1 = A_j - P_j$ | Unitary normalization: $\mathbb{N}1 = 1$ | Mean aggregation, $\mathbb{G}1$ |
| Absolute error: $\mathbb{D}2 = \|A_j - P_j\|$ | Normalization by actuals: $\mathbb{N}2 = A_j^{-c}$ | Median aggregation, $\mathbb{G}2$ |
| Squared error: $\mathbb{D}3 = (A_j - P_j)^2$ | Normalization by variability of actuals: $\mathbb{N}3 = (A_j - \bar{A})^{-c}$ | Geometric mean aggregation, $\mathbb{G}3$ |
| Logarithmic quotient error: $\mathbb{D}4 = \ln(P_j/A_j)$ | Normalization by the sum of actuals and predicted values: $\mathbb{N}4 = (A_j + P_j)^{-c}$ | Sum aggregation, $\mathbb{G}4$ |
| Absolute Log quotient error: $\mathbb{D}5 = \|\ln(P_j/A_j)\|$ | Normalization by maximum (or minimum) value of actuals or predicted: $\mathbb{N}5 = [max(A_j, P_j)]^{-c}$ | |

Note: The values of the variable *c* will be explained in the next section.

**Methods of determining a point distance**, $\mathbb{D}$

It should be noted that the method used to calculate point distance largely determines the overall properties of the performance metric.

In general, point distance can be calculated using any basic mathematical operation: subtraction, addition, multiplication and division (e.g. Deza and Deza, 2016). Commonly, point distances are referred to by the name of the result of the operation, respectively, difference, sum, product, quotient.

Point distances based on subtraction most commonly used in performance metrics and include: error (magnitude of error), $A_j - P_j$; absolute error, $|A_j - P_j|$; and squared error, $(A_j - P_j)^2$. They may be referred to as difference errors (e.g. Willmott, 1985) or just 'errors' as this type by far, the most widely used measure of error in literature.

Subtraction point distances (absolute error and squared error) correspond to the mathematical notions of the Manhattan distance (Taxicab geometry, n.d.) and the Euclidean distance (n.d.), respectively, and their generalization - Minkowski distance (n.d.). More methodological details are provided by McCune, Grace and Urban (2002).

Point distances based on division (similarly to the subtraction distances) include: magnitude of quotient error, $q_j = P_j/A_j$ (referred to as accuracy ratio by Toffalis (2015); absolute quotient error, $|q_j|$; and squared quotient error, $q_j^2$. Note that division point distances are undefined when actual values are zeros.

Kitchenham et al. (2001) introduced quotient error (accuracy ratio) into software development effort forecasting industry, designating it variable z. Although this metric has been studied earlier, as an alternative to subtraction-type errors, in different environments. For example, Olver (1978) used it as an error for basic operations in floating-point arithmetic; also, Törnqvist, Vartia and Vartia (1985) considered this metric as one of relative measures in statistics.

Most commonly, quotient error is used in the form of logarithmic quotient, i.e. $\ln(P_j/A_j)$. Although Tofallis (2015) studied squared quotient error as a loss function in prediction model selection.

Multiplication point distances are more suitable for vector represented data and binary data which are not in scope of this study. Examples can be found in inner product and fidelity groups of metrics in Cha (2007) and Prasath et al (2017): e.g. Inner Product Distance (IPD), Harmonic Mean Distance (HMD) (not to be confused with harmonic mean – aggregation procedure).

To the best of our knowledge, addition point distances were not used in the practical applications in the fields of our interest.

Properties of the commonly used point distances are outlined below.

Error (magnitude of error): $\mathbb{D}1 = A_j - P_j$

The most "natural" method of determining point distance between the actual and predicted values is subtracting one from another. The result of subtraction is a magnitude of error (or just error). Following the currently accepted notation in forecasting, we will be subtracting predicted value from the actual.

Finding the magnitude of error is a straight forward and computationally efficient method. Other methods of determining point distance use the magnitude of error for further processing.

Also, the error is measured with the same units as the data under analysis (variable of interest). It is easily interpretable. In many problems, our business objective or loss function is proportional to the difference between the actual and predicted values (not square or absolute value of this difference, as other point distances imply).

The issue with this method may arise at the aggregation phase, when the positive and negative errors will be cancelling each other. It means that even with large (but having different signs) errors the result of calculating the performance metric may yield zero demonstrating a falsely high accuracy. On another hand, this property of a magnitude of error (showing the direction of error) may convey useful information, e.g. it may be used in analysis to determine whether the forecasting method tends to overestimate or underestimate actual values, i.e. biased. This distance is used in ME, MPE, etc.

Absolute error: $\mathbb{D}2 = |A_j - P_j|$

The idea behind the absolute error is to avoid mutual cancellation of the positive and negative errors. Absolute error has only non-negative values which facilitates aggregation of point distances over the data set.

By the same token, avoiding potential of mutual cancelations has its price – skewness (bias) cannot be determined.

Absolute error preserves the same units of measurement as the data under analysis and gives all individual errors same weights (as compared to squared error). This distance is easily interpretable and when aggregated over a dataset using an arithmetic mean has a meaning of average error.

The use of absolute value might present difficulties in gradient calculation of model parameters (Chai and Draxler, 2014). This distance is used in such popular metrics as MAE, MdAE, etc.

Squared error: $\mathbb{D}3 = (A_j - P_j)^2$

Squared error follows the same idea as the absolute error – avoid negative error values and mutual cancellation of errors.

Due to the square, large errors are emphasized and have relatively greater effect on the value of performance metric (if $e > 1$). At the same time, the effect of relatively small errors ($e < 1$) will be even smaller. Sometimes this property of the squared error is referred to as penalizing extreme errors or being susceptible to outliers. Based on the application, this property may be considered positive or negative. For example, emphasizing large errors may be desirable discriminating measure in evaluating models (Chai and Draxler, 2014).

Squared error has unit measure of squared units of data. This may not be intuitive, e.g. squared dollars. This could be reversed at the aggregation phase by taking square root.

Squared error is acknowledged for its good mathematical properties. It is continuously differentiable which facilitates optimization.

Logarithmic quotient error: $\mathbb{D}4 = \ln(P_j/A_j) = \ln(P_j) - \ln(A_j)$

Logarithmic (Log) quotient error has some useful properties. The error is symmetric (to the change of actual and predicted values in the formula) and dimensionless (e.g. Tornqvist et al, 1985; Tofallis, 2015).

As an example, log quotient distance is used in Median Log Accuracy Ratio (MdLAR) or MdLQ – in author's notation (Morley, 2016; Morley, Brito and Welling, 2018).

Also, quotient distance is used (with normalization which results in non-symmetry) in the Shannon's or entropy-type metrics, e.g. Kullback-Leibler Divergence (KLD) and Jeffreys Divergence (JD) (Kullback and Leibler, 1951; Cha, 2007). Martin et al (2015) found that the KLD-based method in the presence of contaminated noise outperformed the L2-based measure in the global localization of mobile robots experiment.

Absolute Log quotient error: $\mathbb{D}5 = |\ln(P_j/A_j)|$

The intention of taking an absolute value of the log quotient error is to ensure symmetric behaviour of the metric in a sense of possible changing the positions of the predicted and actual values in the formula without altering the result (Morley, Brito and Welling, 2018).

This distance is used in median symmetric accuracy (MdSA) which was developed to enhance certain characteristics of the MAPE (Morley, Brito and Welling, 2018).

Yu et al (2006) proposed two metrics for evaluating air quality models using absolute log quotient error: Mean Normalized Absolute Factor Error (MNAFE) and Mean Normalized Factor Bias (MNFB).

Other point distances

Two more distance metrics have been mentioned in the literature but have not been widely used. First, a time-distance measure of accuracy designed to perform two-dimensional comparisons of time series (Sicherl, 1994; Granger and Jeon, 2003). Second, so called, mean-based measures where error is calculated as $e = \bar{A} - P_j$ for evaluating forecasts against the mean of the underlying process of intermittent demand (Prestwich, et al, 2014). These measures did not gain popularity and have not been included in the final typology.

**Methods of normalization, $\mathbb{N}$**

The main idea behind normalization is to design metrics which can be used to compare multiple series having various dimensions. Most of the normalization methods involve division/multiplication of the point distance by certain parameter. Utilizing operation of mathematical division immediately leads to two properties: first, change of the dimension – often making the metric dimensionless, and second, risks of the denominator to become zero or close to zero and make operation impossible.

It should be emphasized that in our typology normalization is applied to the point distance (each individual error) prior to aggregation phase. There are some metrics with normalization or similar mathematical operations are applied to the aggregated error value. These cases are considered extended metrics.

Unitary normalization: $\mathbb{N}1 = 1$

Unitary normalization – division by one – does not require any calculations and has been included for the generalization purposes. A number of metrics employ unitary normalization, e.g. ME (MBE), MAE (MAD), MdAE, GMAE, MSE. These metrics sustain the dimension of the point distance. So, they are appropriate for analyzing single series, but not useful for comparing multiple series.

Normalization by actuals: $\mathbb{N}2 = A_j^{-c}$

Normalization by actuals involves division of the error by the actual value. For the magnitude of error and absolute error $c = 1$, and for the squared error $c = 2$. Also, for the absolute distance error, absolute actuals are used.

Normalization by actuals is used, for example, in MARE (referred to as MMRE - Mean Magnitude Relative Error – in software effort estimation field, e.g. Jørgensen (2007)).

Commonly, the results are multiplied by 100 to present the ratio as a percentage. Normalization by actuals is used in MPE, MAPE, MdAPE, RMSPE, RMdSPE – often referred to as percentage metrics.

Metrics with normalization by actuals are dimensionless allowing comparison of multiple series.

If actual values are zeros or very close to zeros, the metric cannot be used (undefined due to division by zero). An example of such scenario can be found in predicting intermittent (sporadic) demand (Hyndman, 2006). To avoid a problem of division by zero, Tabataba et al (2017) suggest adding a small value (e.g. the lowest non-zero value of actual data) to $A_j$ in the denominator, calling this algorithm a corrected MAPE (cMAPE).

Obvious analogy with normalization by actuals is normalization by predicted values. This method is mentioned in some papers, e.g. Tofallis (2015), Törnqvist, Vartia and Vartia (1985), but did not become popular in the literature. Fildes and Goodwin (2007) cautioned that inflating predicted values would distort this normalization type. Although the preference of forecaster practitioners towards actuals in denominator is not overwhelming: according to a survey by Green and Tashman (2009) 56% prefer actuals.

Normalization by variability of actuals: $\mathbb{N}3 = (A_j - \bar{A})^{-c}$

Normalization by variability of actuals includes division of the error by the difference between the actual value and mean value of all actuals. For the magnitude of error and absolute error $c = 1$, and for the squared error $c = 2$. Also, for the absolute distance error, absolute actuals are used.

Inclusion of the actuals mean $\bar{A}$ is intended to lower the risk of division-by-zero situations. Actuals mean is implemented in R packages (e.g. in *MLmetrics, metrics, rminer*). In general case, normalization can use an error from a benchmark method (usually naïve forecasting) (Hyndman, 2006).

Normalization by variability of actuals is used in RAE, MRAE, MdRAE, GMRAE, RSE, RRSE – often referred to as relative metrics.

Normalization by the sum of actuals and predicted values: $\mathbb{N}4 = (A_j + P_j)^{-c}$

Normalization by the sum of actual and predicted values involves division of the point distance by the sum of the actuals and predicted values. It was introduced in relation to MAPE. Initial intent behind this type of normalization was to make MAPE symmetric (Makridakis, 1993). However, later it was shown that the objective was not gained – sMAPE (symmetric MAPE) was still asymmetric (Goodwin and Lawton, 1999). At the same time, it seems reasonable to assume that the sum of the actuals and predicted values has less risk to be equal to zero. Several options of this normalization method exist (Symmetric, n.d.). Popular ones use an average of the actuals and predicted values, i.e. $(A_j + P_j)/2$ (Green and Tashman, 2009) or use absolute actual and predicted values.

Normalization by the sum of actual and predicted values is used in sMAPE and sMdAPE – often referred to as 'symmetric' percentage metrics. Also, this normalization is used in FB and FAE (e.g. Yu et al, 2006).

Normalization by maximum (or minimum) value of actuals or predicted: $\mathbb{N}5 = [max(A_j, P_j)]^{-c}$

Normalization by the maximum (or minimum) amount of actuals and predicted values. In the known metrics, $c = 1$. It was introduced in relation to the so called Wave Hedges Distance (e.g. Cha, 2007; Prasath, 2017). This normalization was proved useful in a comparative study of similarity metrics for compressed domain image retrieval (Hatzigiorgaki and Skodras, 2003).

Other normalization methods

Normalization by standard deviation or the difference of the actual and predicted values (as in Mean Normalized Absolute Factor Error - MNAFE) may be used.

All normalization methods described in this subsection have the form of a multiplier (denominator), so the generic formulae for a performance metric can be simplified:

$$\mathbb{m} = \mathop{\mathbb{G}^z}_{j=1,n} \{\mathbb{N}^z \times \mathbb{D}^z(A_j, P_j)\}$$

Although, implementation of more sophisticated methods in the future cannot be excluded.

**Methods of aggregation of point distances over a data set, $\mathbb{G}$**

Aggregation of point distances (in many cases after normalization) over a data set represents the final phase in the calculating primary performance metric.

Mean aggregation, $\mathbb{G}1$

Note that we use the term 'mean' to refer to the 'arithmetic mean'. For any other types of means we add an attribute, e.g. geometric mean. Calculation of the arithmetic mean of the normalized point distances over a data set is the most popular aggregation method (Arithmetic Mean, n.d.). Finding arithmetic average of the observed errors is easy: it involves summing the values of point distances and dividing by the number of elements of the data set. It is also intuitively clear:

the result represents an expected value of the error. The method is used, for example, in MPE, MRAE, MSE, etc. Mean aggregation is sensitive to outliers and skewed data. Refer to Other aggregation methods below for the versions of the mean aggregation intended to overcome issues with asymmetrical distributions of data and extreme values.

Median aggregation, $\mathbb{G}2$

Computation of the median involves listing all point distances in an ordered form by their value (ascending or descending) and finding the number in the centre or the mean of two middle values, if the data set has even number of elements (Median, n.d.).

Opposite to other methods, median method can be called "aggregation" only conditionally: it is not based on some sort of bundling of all point distances of the data set and calculating an output value. The output of this method is one of the existing values of point distances (searched and found through a special procedure).

The median method is more resistant to outliers than the mean (Bakker and Gravemeijer, 2006). On the other hand, there is no clear and easy mathematical formula to describe the method, so theoretical considerations are a cumbersome task (although, computational algorithms present no difficulty and included in most statistical software packages).

The method is used, for example, in MdAE, MdRAE, sMdAPE, etc.

Geometric mean aggregation, $\mathbb{G}3$

The geometric mean is defined as the n-th root of the product of the values of the data set (Geometric Mean, n.d.).

Geometric mean, as a median aggregation, is more robust to outliers than arithmetic mean aggregation (Zhou, Zhou and Mathews, 1999; Fildes, 1992).

As the method includes operations of multiplication and root extraction, the downside of this method is that aggregation is undefined, if the point distances contain negative or zero-value elements.

Makridakis and Hibon (1995) note an advantage of geometric means in interpreting model comparisons: if there are two geometric mean assessments, e.g. 10 and 12, then "the mean absolute errors of the second method are 20% higher than those of the first".

The method is used, for example, in GRMSE (Newbold and Granger, 1974; Fildes, 1992), GMRAE, GMAE, etc.

Sum aggregation, $\mathbb{G}4$

The sum aggregation is just summing point distances to create a simple metric (Sum of absolute differences, n.d.). The method is used, for example, in RAE, SSE, RSE, SAD, etc.

Other aggregation methods

The harmonic mean is calculated as the reciprocal of the arithmetic mean of the reciprocals of the data set (Harmonic mean, n.d.). The harmonic mean (as well as arithmetic and geometric) was known to ancient Greek mathematics since around 500 BC (Heath, 1981). Sometimes all three means referred to as Pythagorean means (n.d.). However, this method is not as popular as the other two means.

The truncated mean (or trimmed mean) is a version of the arithmetic mean. It involves discarding some extreme data points at the high and low end before calculating arithmetic mean on the rest of the data set. This method appears to be more robust to outliers compared to a standard arithmetic mean, but could lead to a biased estimation, if underlying error distribution is not symmetric (Truncated mean, n.d.; Meyer and Venkatu, 2014). Windsorized mean is similar to the truncated mean, except the extreme data points are not discarded but replaced by the next largest (or smallest) values (Winsorized mean, n.d.).

Use of M-estimators is another method to deal with outliers and non-normal distributions which may contaminate arithmetic mean (M-estimator, n.d.). M-estimator is a robust estimator that weights the observations on the basis of their relative distance from the centre of the distribution. Monero et al (2013) proposed using Huber M-estimator to improve performance of the mean absolute percentage error metric. They called this metric Resistant MAPE or R-MAPE.

Similar to the median aggregation, sometimes maximum aggregation is used. It involves searching the maximum value in the point distances. This method is employed in Maximum Absolute Error (MaxAE) (Zhang et al, 2015).

**Visualising Typology**

The developed typology has been visualized using a table format. Table 3 demonstrates that 40 primary metrics have been conveniently ordered and organized by their components shedding light on the properties of the metrics.

Note that for better visualization the table is not comprehensive. It includes only most popular components. For example, MNFB metric is shown on the list in the Appendix 3 with mathematical definitions of metrics by not in the Table because it uses normalizer which is not very common.

**Table 3 Performance metrics (error measures) typology chart**

| Point Distance, $\mathbb{D}$ | Normalization, $\mathbb{N}$ | | | | | Aggregation, $\mathbb{G}$ | |
|---|---|---|---|---|---|---|---|
| | $\mathbb{N}1 = 1$ Unitary | $\mathbb{N}2 = A_j^{-c}$ By Actual Values | $\mathbb{N}3 = (A_j - \bar{A})^{-c}$ By Variability of Actual Values | $\mathbb{N}4 = (A_j + P_j)^{-c}$ By Sum of Actual and Predicted Values | $\mathbb{N}5 = [max(A_j, P_j)]^{-c}$ By Max (or Min) of Actual and Predicted Values | | |
| Error (magnitude of error) $\mathbb{D}1 = A_j - P_j$ | ME (MBE, bias) | MNB  C=1  MPE=100MNB | | FB  C=1 | | $\mathbb{G}1$ | Mean |
| | | | | | | $\mathbb{G}2$ | Median |
| | | | | | | $\mathbb{G}3$ | Geometric Mean |
| | | MD | | | | $\mathbb{G}4$ | Sum |
| Absolute error $\mathbb{D}2 = |A_j - P_j|$ | MAE (MAD) | MARE  C=1  MAPE=100MARE | MRAE  C=1 | FAE  C=1  sMAPE=100FAE | | $\mathbb{G}1$ | Mean |
| | MdAE | MdAPE  C=1 | MdRAE  C=1 | sMdAPE  C=1 | | $\mathbb{G}2$ | Median |
| | GMAE | | GMRAE  C=1 | | | $\mathbb{G}3$ | Geometric Mean |
| | SAD | | RAE  C=1 | CM  C=1 | WHD  C=1  max | $\mathbb{G}4$ | Sum |
| Squared error $\mathbb{D}3 = (A_j - P_j)^2$ | MSE  RMSE = √MSE | MSPE  C=2  RMSPE = √MSPE | | | | $\mathbb{G}1$ | Mean |
| | | MdSPE  C=2  RMdSPE = √MdSPE | | | | $\mathbb{G}2$ | Median |
| | | GRMSE | | | | $\mathbb{G}3$ | Geometric Mean |
| | SSE  ED = √SSE | NCSD  C=1 | RSE  C=2  RRSE = √RSE | SquD  C=1  DivD  C=2 | VSD  C=1  min | $\mathbb{G}4$ | Sum |
| Log quotient error $\mathbb{D}4 = \ln(P_j/A_j) = \ln(P_j) - \ln(A_j)$ | | | | | | $\mathbb{G}1$ | Mean |
| | MdLAR | | | | | $\mathbb{G}2$ | Median |
| | | | | | | $\mathbb{G}3$ | Geometric Mean |
| | | KLD  C=-1 | | | | $\mathbb{G}4$ | Sum |
| Absolute Log quotient error $\mathbb{D}5 = |\ln(P_j/A_j)|$ | MNAFE | | | | | $\mathbb{G}1$ | Mean |
| | MdSA | | | | | $\mathbb{G}2$ | Median |
| | | | | | | $\mathbb{G}3$ | Geometric Mean |
| | | | | | | $\mathbb{G}4$ | Sum |

Discussion

The paper provided an overview of a wide range of performance metrics used in machine learning regression, forecasting and prognostics. A comparison of prior metrics classifications and their limitations was conducted. Prior typologies are based on a one-level ("flat") structure with 5-9 categories which made it difficult to organize multiple metrics without overlappings. Our typology suggests two levels with a detailed typology of primary metrics which allows incorporating more metrics than it was possible with prior classifications. Suggested typology has been shown to cover most of the commonly used primary metrics – total of over 40.

Also, prior typologies group together metrics with significant differences. For example, Hyndman's classification arranges together metrics based on different errors – absolute and squared – although these metrics have considerably different properties.

Finally, prior typologies operate with metrics taken as complete structures without going deeper into the metric construct. Our typology defines metrics components which determine metrics' properties.

Suggested in this paper generic formula for primary performance metrics is more comprehensive than used by Willmott & Matsuura (2005), as their definition can be applied only to metrics with mean-averaging type of error aggregation.

The developed typology can inform metric selection process decision making by structuring performance metrics considerations (point distance, normalization and aggregation phases) and focusing on the key properties of the components chosen. For example, if the business or research need is to emphasize outliers, squared error and arithmetic mean should be used. However, if the business requirement is to isolate outliers, then selection of absolute error and geometric mean is desirable. In other words, the use of this typology turns selection of a metric from a browsing exercise over dozens of metrics into a straightforward process of identifying point distance, normalization and aggregation methods that fit the purpose of the task.

The benefits of the developed typology, outlined above, are also applicable to the process of facilitating creation of new metrics. It should be noted that this study have not revealed recently conceived types of point distances, normalizers or aggregators - all of them existed for a while. Suggested visualization table can be used as a tool for creating new metrics by consciously choosing blank cells in the chart (an analogy with the Periodic Table of the chemical elements).

Assumptions and limitations

It has been shown that the typology developed in this paper can be applied to a wide variety of commonly used performance metrics. However, not all existing metrics can directly match the typology. For example, mean arctangent absolute percentage error (MAAPE) proposed by Kim and Kim (2016). MAAPE is a modification of MAPE which involves taking arctangent of the absolute error normalized by the actual values.

Our approach in this study is conceptual. We are not empirically comparing various metrics, but rather consider their qualitative properties.

We focus on numerical data. Metrics for evaluating categorical, ordinal, binary types of data are not in scope. Finally, within machine learning metrics we consider only metrics used in regression (not classification or clustering).

**Concluding Remarks**

The importance and timeliness of the paper is determined by the increased interest of researchers and practitioners to improving evaluation results in machine learning regression, forecasting and prognostics. The paper overviewed multiple performance metrics and conducted a comparison of prior metrics classifications.

The main findings and results of the study include the following.

The paper proposed metrics framework, which includes four (4) categories: primary metrics, extended metrics, composite metrics and hybrid sets of metrics.

The paper identified three (3) key components (dimensions) that determine the structure and properties of primary metrics: method of determining point distance, method of normalization, method of aggregation of point distances over a data set.

The paper proposed a new primary metrics typology designed around the key metrics components. The suggested typology has been shown to cover most of the commonly used primary metrics – total of over 40.

A new generic mathematic formula for primary performance metrics has been proposed which implies sequential determining the point distance between the actual and predicted values, normalizing it and then aggregating results over a complete data set.

Typology visualization chart has been designed which can be used as a tool for assessing existing and creating new metrics.

The main contribution of this paper is in ordering knowledge of performance metrics and enhancing understanding of their structure and properties by proposing a new typology, generic primary metrics mathematic formula and a visualization chart.

The practical significance of the paper is in the fact that the presented findings can be used to facilitate teaching performance metrics to university students, expedite metrics selection process for practitioners and streamline new metrics development for academics.

Two future research opportunities can be conceived from the results of this paper. First, following the approach taken in this paper to model and analyze primary metrics, to continue conceptual research into the properties of the other metrics categories identified in this paper, namely: extended metrics, composite metrics and hybrid sets of metrics. Second, start an empirical study of the metrics, using R Studio or Azure Machine Learning Studio, to find associations between the conceptual properties of primary metrics and their "numerical" behavior in a wide spectrum of data characteristics and business or research requirements.

Willmott, C. J., Ackleson, S. G., Davis, R. E., Feddema, J. J., Klink, K. M., Legates, D. R., O'Donnell, J., & Rowe, C. M. (1985). Statistics for the evaluation and comparison of models. *Journal of Geophysical Research: Oceans, 90*(C5), 8995-9005.

Willmott, C. J., & Matsuura, K. (2005). Advantages of the mean absolute error (MAE) over the root mean square error (RMSE) in assessing average model performance. *Climate research*, *30*(1), 79-82.

Willmott, C. J., Matsuura, K., & Robeson, S. M. (2009). Ambiguities inherent in sums-of-squares-based error statistics. *Atmospheric Environment*, *43*(3), 749-752.

Winsorized mean. (n.d.). In *Wikipedia*. Retrieved July 19, 2018, from https://en.wikipedia.org/wiki/Winsorized_mean

Yu, S., Eder, B., Dennis, R., Chu, S. H., & Schwartz, S. E. (2006). New unbiased symmetric metrics for evaluation of air quality models. *Atmospheric Science Letters, 7*(1), 26-34. DOI: 10.1002/asl.125

Zhang, J., Florita, A., Hodge, B. M., Lu, S., Hamann, H. F., Banunarayanan, V., & Brockway, A. M. (2015). A suite of metrics for assessing the performance of solar power forecasting. *Solar Energy, 111*, 157-175. http://dx.doi.org/10.1016/j.solener.2014.10.016

Zhou, Q. H., Zhou, Q. N., & Mathews, J. D. (1999). Arithmetic average, geometric average, and ranking: Application to incoherent scatter radar data processing. *Radio Science, 34*(5), 1227-1237.

**Appendix 1. List of metrics abbreviations**

| Metric Abbreviation | Metric Name |
|---|---|
| CM | Canberra Metric |
| CoD | Coefficient of Determination |
| CVRMSE | Coefficient of variation of the RMSE |
| DivD | Divergence Distance |
| ED | Euclidean Distance (L2-norm) |
| FAE | Fractional absolute error |
| FB | Fractional Bias |
| GMAE | Geometric Mean Absolute Error |
| GMRAE | Geometric Mean Relative Absolute Error |
| GRMSE | Geometric Root Mean Squared Error |
| HMD | Harmonic Mean Distance (not to be confused with harmonic mean – aggregation procedure) |
| IPD | Inner Product Distance |
| JD | Jeffreys Divergence |
| KLD | Kullback-Leibler Divergence |
| LMR | Log Mean Squared Error Ratio |
| MAAPE | Mean Arctangent Absolute Percentage Error |
| MAD | Mean Absolute Deviation |
| MAE | Mean Absolute Error |
| MAGE | Mean Absolute Gross Error |
| MAPE | Mean Absolute Percentage Error |
| MARE | Mean Absolute Relative Error |
| MASE | Mean Absolute Scaled Error |
| MaxAE | Maximum Absolute Error |
| MBE | Mean Bias Error |
| MCD | Mean Character Difference |
| MD | Manhattan Distance |
| MdAE | Median Absolute Error |
| MdAPE | Median Absolute Percentage Error |
| MdASE | Median Absolute Scaled Error |
| MdLAR | Median Log Accuracy Ratio |
| MdRAE | Median Relative Absolute Error |
| MdSA | Median Symmetric Accuracy |
| MdSPE | Median Square Percentage Error |
| ME | Mean Error |
| MMRE | Mean Magnitude Relative Error |
| MNAFE | Mean Normalized Absolute Factor Error |
| MNB | Mean Normalized Bias |
| MNFB | Mean Normalized Factor Bias |
| MPE | Mean Percentage Error |
| MRAE | Mean Relative Absolute Error |
| MSE | Mean Squared Error |
| MSPE | Mean Square Percentage Error |
| NCSD | Neyman Chi-Square Distance |
| NMSE | Normalized Mean Squared Error (normalized by variance) |
| NRMSE_m | Normalized Root Mean Squared Error (normalized by the mean of actual data) |
| NRMSE_mm | Normalized Root Mean Squared Error (normalized by the difference between maximum and minimum actual data) |
| NRMSE_sd | Normalized Root Mean Squared Error (normalized by the standard deviation of the actual data) |
| RAE | Relative Absolute Error |
| RelRMSE | Relative Root Mean Square Error |
| RMAE | Relative Mean Absolute Error |
| RMdSPE | Root Median Square Percentage Error |
| RMSE | Root Mean Squared Error |
| RMSPE | Root Mean Square Percentage Error |
| RMSSE | Root Mean Squared Scaled Error |
| RRSE | Root Relative Squared Error |
| RSE | Relative Squared Error |
| SAD | Sum of absolute differences |
| sMAPE | Symmetric Mean Absolute Percentage Error |
| SMdAPE | Symmetric Median Absolute Percentage Error |
| SquD | Squared Chi-square Distance |

SSE             Sum of Squared Error (Squared Euclidean)
VSD             Vicis Symmetric Distance
WHD             Wave Hedges Distance

## Appendix 2. Metrics mathematical definitions

**Note 1.** Legend: $A_j$ – actual values; $\bar{A}$ – the mean of the actual values; $P_j$ – predicted values; $e_j = A_j - P_j$ – error; $n$ – size of the data set

**Note 2.** Metrics are listed according to the categories they belong to, i.e. primary, extended, composite, hybrid sets.; and within categories – by type of error.

| Metric Abbreviation | Metric Name (alternative names are given in brackets) | Metric Formula |
|---|---|---|
| **PRIMARY METRICS** | | |
| | **Error (magnitude of error):** $\mathbb{D}1 = A_j - P_j = e_j$ | |
| ME | Mean Error (Mean Bias Error) | $ME = \dfrac{1}{n} \sum\limits_{j=1}^{n} e_j$ |
| MNB | Mean Normalized Bias | $MNB = \dfrac{1}{n} \sum\limits_{j}^{n} \dfrac{e_j}{A_j}$ |
| MPE | Mean Percentage Error | $MPE = \dfrac{100}{n} \sum\limits_{j}^{n} \dfrac{e_j}{A_j}$ |
| FB | Fractional Bias | $FB = \dfrac{1}{n} \sum\limits_{j}^{n} \dfrac{2 * e_j}{A_j + P_j}$ |
| MD | Manhattan Distance (City Block, $L_1$-norm, Taxicab norm) | $MD = \sum\limits_{j=1}^{n} e_j$ |
| | | |
| | **Absolute error:** $\mathbb{D}2 = |A_j - P_j| = |e_j|$ | |
| MAE | Mean Absolute Error (Mean Absolute Deviation – MAD; Mean Absolute Gross error; | $MAE = \dfrac{1}{n} \sum\limits_{j=1}^{n} |e_j|$ |

| | Mean Character Difference – MCD; Average Manhattan; Gower) | |
|---|---|---|
| MdAE | Median Absolute Error | $MdAE = \underset{j=1,n}{Md}(|e_j|)$ |
| MaxAE | Maximum Absolute Error | $MaxAE = \underset{j=1,n}{max}(|e_j|)$ |
| MARE | Mean Absolute Relative Error (Mean Magnitude Relative Error – MMRE) | $MARE = \frac{1}{n}\sum_{j}^{n}\frac{|e_j|}{|A_j|}$ |
| MAPE | Mean Absolute Percentage Error | $MAPE = \frac{100}{n}\sum_{j}^{n}\frac{|e_j|}{|A_j|}$ |
| MdAPE | Median Absolute Percentage Error | $MdAPE = 100 * \underset{j=1,n}{Md}(\frac{|e_j|}{|A_j|})$ |
| RAE | Relative Absolute Error | $RAE = \sum_{j=1}^{n}\frac{|e_j|}{|A_j - \bar{A}|}$ |
| MRAE | Mean Relative Absolute Error | $MRAE = \frac{1}{n}\sum_{j=1}^{n}\frac{|e_j|}{|A_j - \bar{A}|}$ |
| GMAE | Geometric Mean Absolute Error | $GMAE = \sqrt[n]{\prod_{j=1}^{n}|e_j|}$ |
| SAD | Sum of Absolute Differences | $SAD = \sum_{j=1}^{n}|e_j|$ |

| GMRAE | Geometric Mean Relative Absolute Error | $$GMRAE = exp\left(\frac{1}{n}\sum_{j=1}^{n}\ln\left(\frac{|e_j|}{|A_j - \bar{A}|}\right)\right)$$ or $$= \sqrt[n]{\prod_{j=1}^{n}\left(\frac{|e_j|}{|A_j - \bar{A}|}\right)}$$ |
|---|---|---|
| MdRAE | Median Relative Absolute Error | $$MdRAE = \underset{j=1,n}{Md}\left(\frac{|e_j|}{|A_j - \bar{A}|}\right)$$ |
| WHD | Wave Hedges Distance | $$WHD = \sum_{j=1}^{n}\frac{|e_j|}{max(A_j, P_j)}$$ |
| FAE | Fractional absolute error | $$FAE = \frac{1}{n}\sum_{j}^{n}\frac{2*|e_j|}{|A_j| + |P_j|}$$ |
| sMAPE | Symmetric Mean Absolute Percentage Error | $$sMAPE = \frac{100}{n}\sum_{j}^{n}\frac{2*|e_j|}{|A_j| + |P_j|}$$ |
| SMdAPE | Symmetric Median Absolute Percentage Error | $$SMdAPE = 100 * \underset{j=1,n}{Md}\left(\frac{2*|e_j|}{|A_j| + |P_j|}\right)$$ |
| CM | Canberra Metric | $$CM = \sum_{j}^{n}\frac{|e_j|}{A_j + P_j}$$ |
| | | |
| **Squared error:** $\mathbb{D}3 = (A_j - P_j)^2 = e_j^2$ | | |
| MSE | Mean Squared Error | $$MSE = \frac{1}{n}\sum_{j=1}^{n}e_j^2$$ |

| RMSE | Root Mean Squared Error (Average Distance) | $$RMSE = \sqrt{\frac{\sum_{j=1}^{n} e_j^2}{n}}$$ or $$RMSE = \sqrt{MSE}$$ |
|---|---|---|
| SSE | Sum of Squared Error (Squared Euclidean) | $$SSE = \sum_{j=1}^{n} e_j^2$$ |
| ED | Euclidean Distance ($L_2$-norm) | $$ED = \sqrt{\sum_{j=1}^{n} e_j^2}$$ or $$ED = \sqrt{SSE}$$ |
| VSD | Vicis Symmetric Distance | $$VSD = \sum_{j=1}^{n} \frac{e_j^2}{min(A_j, P_j)}$$ |
| NCSD | Neyman Chi-Square Distance | $$NCSD = \sum_{j=1}^{n} \frac{e_j^2}{A_j}$$ |
| SquD | Squared Chi-square Distance | $$SquD = \sum_{j=1}^{n} \frac{e_j^2}{A_j + P_j}$$ |
| DivD | Divergence Distance | $$DivD = 2\sum_{j=1}^{n} \frac{e_j^2}{(A_j + P_j)^2}$$ |
| RSE | Relative Squared Error | $$RSE = \sum_{j=1}^{n} \frac{e_j^2}{(A_j - \bar{A})^2}$$ |

| RRSE | Root Relative Squared Error | $$RRSE = \sqrt{\sum_{j=1}^{n} \frac{e_j^2}{(A_j - \bar{A})^2}}$$ |
|---|---|---|
| GRMSE | Geometric Root Mean Squared Error | $$GRMSE = \sqrt[2n]{\prod_{j=1}^{n} e_j^2}$$ |
| MSPE | Mean Square Percentage Error | $$MSPE = \frac{100}{n} \sum_{j}^{n} \left(\frac{|e_j|}{|A_j|}\right)^2$$ |
| MdSPE | Median Square Percentage Error | $$MdSPE = 100 * \underset{j=1,n}{Md} \left(\frac{|e_j|}{|A_j|}\right)^2$$ |
| RMSPE | Root Mean Square Percentage Error | $$RMSPE = \sqrt{\frac{100}{n} \sum_{j}^{n} \left(\frac{|e_j|}{|A_j|}\right)^2}$$ |
| RMdSPE | Root Median Square Percentage Error | $$RMdSPE = \sqrt{100 * \underset{j=1,n}{Md} \left(\frac{|e_j|}{|A_j|}\right)^2}$$ |
| | | |
| **Logarithmic quotient error: $\mathbb{D}4 = \ln(P_j/A_j) = \ln(P_j) - \ln(A_j)$** | | |
| MdLAR | Median Log Accuracy Ratio | $$MdLAR = \underset{j=1,n}{Md} (\ln(P_j/A_j))$$ |
| KLD | Kullback-Leibler Divergence | $$KLD = \sum_{j=1}^{n} P_j \ln(P_j/A_j)$$ |
| JD | Jeffreys Divergence | $$JD = \sum_{j=1}^{n} (P_j - A_j) \ln(P_j/A_j)$$ |

|  | Absolute Log quotient error: $\mathbb{D}5 = |\ln(P_j/A_j)|$ | |
|---|---|---|
| MNAFE | Mean Normalized Absolute Factor Error | $MNAFE = \frac{1}{n}\sum_{j=1}^{n}|\exp(|\ln(\frac{P_j}{A_j})|) - 1|$ |
| MNFB | Mean Normalized Factor Bias | $MNFB = \frac{1}{n}\sum_{j=1}^{n}\frac{P_j - A_j}{|P_j - A_j|}[\exp(|\ln(\frac{P_j}{A_j})|) - 1]$ |
| MdSA | Median Symmetric Accuracy | $MdSA = 100(\exp(\underset{j=1,n}{Md}(|\ln(P_j/A_j)|))-1)$ |
|  |  |  |
| **EXTENDED METRICS** | | |
| NRMSE_m | Normalized Root Mean Squared Error (normalized by the mean of actual data) (CVRMSE - coefficient of variation of the RMSE) | $NRMSE\_m = \frac{RMSE}{\bar{A}}$ |
| NRMSE_sd | Normalized Root Mean Squared Error (normalized by the standard deviation of the actual data) | $NRMSE\_sd = \frac{RMSE}{sd}$ |
| NRMSE_mm | Normalized Root Mean Squared Error (normalized by the difference between maximum and minimum actual data) | $NRMSE\_mm = \frac{RMSE}{maxA - minA}$ |
| NMSE | Normalized Mean Squared Error (normalized by variance) | $NMSE = \frac{MSE}{\sigma^2}$ |
|  |  |  |

| | COMPOSITE METRICS | |
|---|---|---|
| RMAE | Relative Mean Absolute Error | $RMAE = MAE/MAE \text{ in-sample}$ |
| RelRMSE | Relative Root Mean Square Error | $RelRMSE = RMSE/RMSE \text{ in-sample}$ |
| LMR | Log Mean Squared Error Ratio | $LMR = \log(RMSE/RMSE \text{ in-sample})$ |
| CoD | Coefficient of Determination | $CoD = 1 - \dfrac{\sum_{j=1}^{n}(P_j - A_j)^2}{\sum_{j=1}^{n}(A_j - \bar{A})^2}$ |
| MASE | Mean Absolute Scaled Error | $MASE = MAE/MAE \text{ in-sample, naïve}$ <br><br> $MASE = MAE/Q,$ <br> where <br> $Q = \dfrac{1}{n-1}\sum_{j=2}^{n}|A_j - A_{j-1}|$ |

**Appendix 3. Performance metrics alternative mathematical definitions**

| Metric Abbreviation | Metric Name (alternative names are given in brackets) | Metric Formula |
|---|---|---|
| RAE | Relative Absolute Error | Option 1 <br><br> $RAE = \sum_{j=1}^{n}\dfrac{|e_j|}{|A_j - \bar{A}|}$ <br><br> Option 2 <br><br> $RAE = \dfrac{\sum_{j=1}^{n}|e_j|}{\sum_{j=1}^{n}|A_j - \bar{A}|}$ |
| MRAE | Mean Relative Absolute Error | Option 1 |
| | | $$MRAE = \frac{1}{n} \sum_{j=1}^{n} \frac{|e_j|}{|A_j - \bar{A}|}$$ Option 2 $$MRAE = \frac{\sum_{j=1}^{n} |e_j|}{n \sum_{j=1}^{n} |A_j - \bar{A}|}$$ |
| RSE | Relative Squared Error | Option 1 $$RSE = \sum_{j=1}^{n} \frac{e_j^2}{(A_j - \bar{A})^2}$$ Option 2 $$RSE = \frac{\sum_{j=1}^{n} e_j^2}{\sum_{j=1}^{n} (A_j - \bar{A})^2}$$ |
| RRSE | Root Relative Squared Error | Option 1 $$RRSE = \sqrt{\sum_{j=1}^{n} \frac{e_j^2}{(A_j - \bar{A})^2}}$$ Option 2 $$RRSE = \sqrt{\frac{\sum_{j=1}^{n} e_j^2}{\sum_{j=1}^{n} (A_j - \bar{A})^2}}$$ |